\documentclass[aps,prd,twocolumn,a4paper,showpacs,superscriptaddress,nofootinbib,floatfix]{revtex4}

\usepackage{latexsym}
\usepackage{graphics}
\usepackage{graphicx}
\usepackage{bm}
\usepackage{amsmath}
\usepackage{amsfonts}
\usepackage{amssymb}
\usepackage{epsfig,array}

\newcommand{\beq}{\begin{equation}}
\newcommand{\eeq}{\end{equation}}
\newcommand{\beqa}{\begin{eqnarray}}
\newcommand{\eeqa}{\end{eqnarray}}

\newcommand{\sla}[1]%
        {\kern .25em\raise.18ex\hbox{$/$}\kern-.75em #1}
\newcommand{\mybar}[1]%
        {\kern 0.8pt\overline{\kern -0.8pt#1\kern -0.8pt}\kern 0.8pt}
\newcommand{\gsim}{\, \raisebox{-0.8ex}{$\stackrel{\textstyle >}{\sim}$ }}
\newcommand{\lsim}{\, \, \raisebox{-0.8ex}{$\stackrel{\textstyle <}{\sim}$ }}

\def\beq{\begin{equation}}
\def\eeq{\end{equation}}
\def\bea{\begin{array}}
\def\eea{\end{array}}
\def\be{\begin{equation}}
\def\ee{\end{equation}}
\def\ba{\begin{eqnarray}}
\def\ea{\end{eqnarray}}

\def\to{\rightarrow}

\def\[{\left[}
\def\]{\right]}
\def\({\left(}
\def\){\right)}

\def\sm0{{\widetilde{m}_0}}

\def\U1em{{U(1)_{\rm em}}}
\def\to{\rightarrow}

\def\sq2{\sqrt{2}}
\def\gaga{\gamma\gamma}

\def\ee{e^+e^-}

\def\End{\end{document}}

\def\fsl#1{\setbox0=\hbox{$#1$}                 
   \dimen0=\wd0                                 
   \setbox1=\hbox{/} \dimen1=\wd1               
   \ifdim\dimen0>\dimen1                        
      \rlap{\hbox to \dimen0{\hfil/\hfil}}      
      #1                                        
   \else                                        
      \rlap{\hbox to \dimen1{\hfil$#1$\hfil}}   
      /                                         
   \fi}

\newcommand{\bd}{\begin{displaymath}}
\newcommand{\ed}{\end{displaymath}}

\def\lsim{\raise0.3ex\hbox{$\;<$\kern-0.75em\raise-1.1ex
\hbox{$\sim\;$}}}
\def\gsim{\raise0.3ex\hbox{$\;>$\kern-0.75em\raise-1.1ex
\hbox{$\sim\;$}}}
\def\simlt{\mathrel{\lower2.5pt\vbox{\lineskip=0pt\baselineskip=0pt
           \hbox{$<$}\hbox{$\sim$}}}}
\def\simgt{\mathrel{\lower2.5pt\vbox{\lineskip=0pt\baselineskip=0pt
           \hbox{$>$}\hbox{$\sim$}}}}
\def\unity{{\hbox{1\kern-.8mm l}}}

\def\16p{16\pi^2}

\def\al{\alpha}

\renewcommand{\to}{\rightarrow}

\def\bc{\begin{center}}
\def\ec{\end{center}}

\begin{document}

\title{
Supersymmetric Higgs mediated lepton flavor violation
at a Photon Collider
}

\author{\textsc{M.~Cannoni}}
\affiliation{Universit\`a di Perugia, Dipartimento di Fisica,
Via A.~Pascoli, I-06123, Perugia, Italy}
\affiliation{Istituto Nazionale di Fisica Nucleare, Sezione di Perugia, Via A.~Pascoli, 06123 Perugia, Italy}

\author{\textsc{O.~Panella}}
\affiliation{Istituto Nazionale di Fisica Nucleare, Sezione di Perugia, Via A.~Pascoli, 06123 Perugia, Italy}

\date{December 15, 2008}

\begin{abstract}
We study a new signature of lepton flavor violation (LFV) at the Photon Collider (PC) within Supersymmetric (SUSY) theories. We
consider  the minimal supersymmetric standard model within a large $\tan\beta$ scenario with all superpartner masses in the ${\cal O}$(TeV) while the heavy Higgs bosons masses lie below the TeV and develop sizable loop induced LFV
couplings to the leptons. We consider a photon collider based on an $e^+e^-$ linear collider with $\sqrt{s}=800$ GeV with the parameters of the TESLA proposal and  show that, with the expected integrated $\gamma\gamma$-luminosity $L_{\gamma\gamma}=200\div 500$ fb$^{-1}$, the ``$\mu\tau$ fusion'' mechanism is the dominant channel for the
process $\gaga\to\mu\tau b\bar{b}$
providing detailed analytical and numerical studies of the signal and  backgrounds.
We impose on the parameter space
present direct and indirect constraints from $B$ physics and rare LFV $\tau$-decays and find that the LFV signal
can be probed for masses of the heavy neutral Higgs bosons
$A,H$ from $300$ GeV up to the kinematical limit $\simeq 600$ GeV for 30$\leq\tan\beta\leq$60.

\end{abstract}


\maketitle

\section{Introduction}
\label{sec1}

There is an emergent consensus in the physics community that the
next  complementary step to the Cern large hadron collider (LHC)
will be the International Linear Collider (ILC) which will collide
$e^+ e^-$ beams with a center of mass energy in the range $2E_e
=500-1000$ GeV~\cite{Brau:2007zza,ILC}. It is also well known, since the pioneering work of the Novosibirsk school, that such a linear collider could offer the possibility of working in the so called $e\gamma$ or $\gamma\gamma$ modes thus realizing a very high energy  photon collider
(PC) with polarized photon beams~\cite{ginzburg}.

A vast literature is already available on
the ILC and the PC potentialities for the discovery and precision
measurement of the properties of the Higgs boson, the missing piece
of the Standard Model (SM) of electro-weak interactions. Moreover,
the properties of extended Higgs sectors of many well motivated
theories beyond the SM have been widely considered as well. Among
these, the Minimal Supersymmetric Standard Model (MSSM) has received
particular attention, for reviews see~\cite{djouadi1,djouadi2}.

However, the
MSSM (like the SM) does not provide any explanation for the neutrino
masses and mixing. In order to accomplish this task, the seesaw
mechanism is usually implemented in the MSSM. Hence we are led to
study a MSSM with right handed neutrinos ($\nu$-MSSM). Compared to the
MSSM, the main novelty in the $\nu$-MSSM is the presence of lepton
flavor violation (LFV). LFV effects arise both in the
gauge interactions~\cite{borzumati} (through lepton-slepton-gaugino couplings)
and in the Yukawa interactions~\cite{bkl}.
In particular, LFV Yukawa interactions are greatly enhanced at large $\tan\beta$, and give
the possibility of detecting LFV decays of the Higgs bosons at LHC~\cite{brignole1,moretti} and ILC in the $e^+e^-$ mode~\cite{kanemura1}.
In Refs.\cite{Cannoni:2003my,Cannoni:2005gy,Cannoni:2006hv} loop level lepton flavor violating processes such as $e^+e^- \to e^+\ell^-$ ($\ell=\mu\tau$),
and $\gamma\gamma\to \ell_i\ell_j$ ($\ell_i \neq \ell_j$),
which are potentially striking signatures of LFV, were studied in detail.

In this work we extend these previous studies and discuss a new mechanism of lepton flavor violation at the photon collider via the Higgs mediated ($H,A$) process:
\begin{equation}
\gamma \gamma \to \mu \tau b \bar{b}
\end{equation}
in a  scenario of large $\tan\beta$ where all the super-partner masses are $\cal{O}$(TeV) and
the heavy Higgs bosons ($A,H$) have instead masses below the TeV and develop sizable
loop induced LFV couplings to the SM leptons.

\begin{figure}[t!]
\begin{center}
\includegraphics*[scale=0.6]{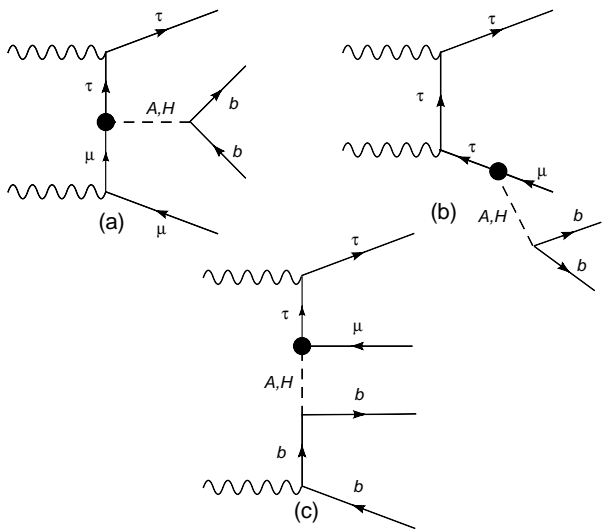}
\caption{Diagrams for the process $\gamma\gamma \to \mu\tau
b\bar{b}$: the topology (a) is the one we call $\mu\tau$ fusion. The
black blob represents the loop induced LFV coupling treated as an
effective vertex. }
\label{diagrams}
\end{center}
\end{figure}

In photon-photon collisions the main production mechanisms for the Higgs bosons
are $\gamma\gamma$ fusion~\cite{djouadi1} (and references therein) and $\tau\tau$ fusion~\cite{choi}.
In the first case, the Higgs is produced as an s-channel resonance through a loop involving the exchange of massive charged particles.
In the $\tau\tau$ fusion process $\gamma\gamma \to \tau\tau b\bar{b}$,
the Higgs is produced in the s-channel with a $\tau\tau$ pair and
can be detected with its decay mode $b\bar{b}$.
We show that the main LFV process is the $\mu\tau$ fusion to the Higgs,
see Fig. (\ref{diagrams}a), which dominate the $\gaga$ fusion, with large cross section over large
portion of the parameter space. The signal from the $\mu\tau$ fusion
$\gaga \to \mu\tau b\bar{b}$ consists of a $\mu\tau$ pair plus $b\bar{b}$ jets
from the Higgs decay, allowing the possibility to detect and reconstruct the Higgs
through its main decay channel and to measure, at the same time, the size of LFV
couplings. In Ref.~\cite{choi} it is  shown that the lepton flavor
conserving (LFC) channel $\gaga \to \tau\tau b\bar{b}$ (``$\tau\tau$ fusion'')
allows to measure $\tan \beta$ with a precision which is better than $10\%$ at large $\tan\beta$.

The plan of the paper is the following.
In Section~\ref{sec2} we
review  the theory of lepton flavor violation related to the MSSM Higgs bosons and
discuss their properties within  our scenario.
Section~\ref{sec3} is devoted to the analytical
evaluation of the signal cross sections at the photon collider; in Section~\ref{signalback}
we present numerical simulations of both the signal and the background.
The correlations among the signal at the PC and the constraints
imposed by $B$ physics and the non observation of SUSY  particles and
lepton flavor violating rare $\tau$-decays is discussed in Section~\ref{sec5}.
Finally, we give a summary and the conclusions in Section~\ref{sec6}.

\section{Higgs LFV in SUSY}
\label{sec2}
Within a SUSY framework, LFV effects originate from any misalignment
between fermion and sfermion mass eigenstates.
In particular, if the light neutrino masses are obtained via a
see-saw mechanism, the radiatively induced off-diagonal (LFV) entries in the slepton
mass matrix $(m^2_{\tilde{L}})_{ij}$ are given by~\cite{borzumati,hisano1,hisano2}:
\begin{equation}
\label{mLfromseesaw}
(m^2_{L})_{i\neq j} \approx
- \frac{3m^2_0}{8\pi^2} (Y_{\nu} Y_{\nu}^\dagger)_{i\neq j}
\ln \left(\frac{M_X}{M_{R}} \right)\,,
\end{equation}
where $M_X$ denotes the scale of SUSY-breaking mediation, $M_R$
the scale of the heavy right-handed neutrinos masses, $m_0$
the universal supersymmetry breaking scalar mass and
$Y_{\nu}$ the Yukawa couplings between left- and right-handed neutrinos (the
potentially large sources of LFV).
Since the see-saw equation $m_\nu = - Y_\nu \hat{M}^{-1}_R Y_\nu^T
\langle H_u \rangle^2$, with $\langle H_u \rangle$ is the
vacuum expectation value of the up-type Higgs,
allows large $(Y_\nu
Y_\nu^\dagger)$ entries, sizable effects can stem from
this running.
The determination of $(m^2_{\tilde{L}})_{i\neq j}$ would imply a
complete knowledge of the neutrino Yukawa matrix $(Y_\nu)_{ij}$,
which is not possible even if all the low-energy observables from
the neutrino sector were known~\cite{masiero1}. As a result,
the predictions of
leptonic flavor changing neutral current effects will remain undetermined even in the very
optimistic situation where all the relevant New Physics masses were
measured at the LHC.
More stable predictions can be obtained embedding the SUSY model within
a Grand Unified Theory (GUT) where the see-saw mechanism can naturally
arise (such as $SO(10)$)~\cite{masiero2}. In this case the GUT symmetry allows us
to obtain some hints about the unknown neutrino Yukawa matrix $Y_{\nu}$.

There exist two different classes of LFV contributions to rare
decays:
gauge-mediated LFV effects through the exchange of gauginos and sleptons
~\cite{borzumati,hisano1,hisano2}
and Higgs-mediated LFV effects through effective non-holomorphic
Yukawa interactions for quarks and leptons
~\cite{bkq,bkl}:
once a source of LFV is given in the slepton mass matrix, for example
Eq. (\ref{mLfromseesaw}) in MSSM with the see saw mechanism,
LFV Yukawa coupling of the type $\bar{L}_R^i L_L^j H_u^*$ are induced at loop
level and become particularly sizable at large $\tan\beta$.

In the mass-eigenstate basis for both leptons and Higgs bosons,
the effective flavor-violating Yukawa interactions are described by the
lagrangian:
\beqa
\mathcal{-L}&\simeq&(2G_{F}^2)^{\frac{1}{4}}\frac{m_{l_i}} {c^2_{\beta}}
\left(\Delta^{ij}_{L}\overline{l}^i_R l^j_L+\Delta^{ij}_{R}
\overline{l}^i_L l^j_R \right)\nonumber\\
&\times&\left(c_{\beta-\alpha}h-s_{\beta-\alpha}H-iA \right) +h.c.
\nonumber\\
&+&
(8G_{F}^2)^{\frac{1}{4}}\frac{m_{l_i}} {c^2_{\beta}}
\left(\Delta^{ij}_{L}\overline{l}^i_R \nu^j_L+\Delta^{ij}_{R}\nu^i_L\overline{l}^j_R\right)
H^{\pm} + h.c.\nonumber\\
\label{lagrangian}
\eeqa
where $\alpha$ is the mixing angle between the CP-even Higgs bosons $h$ and
$H$, $A$ is the physical CP-odd boson,
$H^{\pm}$ are the physical
charged Higgs-bosons
and  we adopt the notation
$(c_{\theta},s_{\theta},t_{\theta})\! =\!(\cos \theta,\sin \theta,\tan \theta)$. We note that in Eq.~(\ref{lagrangian}) $i,j$ are flavor indices that in the following are understood to be different ($i\neq j$).
\begin{figure*}[ht!]
\begin{center}
\includegraphics*[scale=0.625]{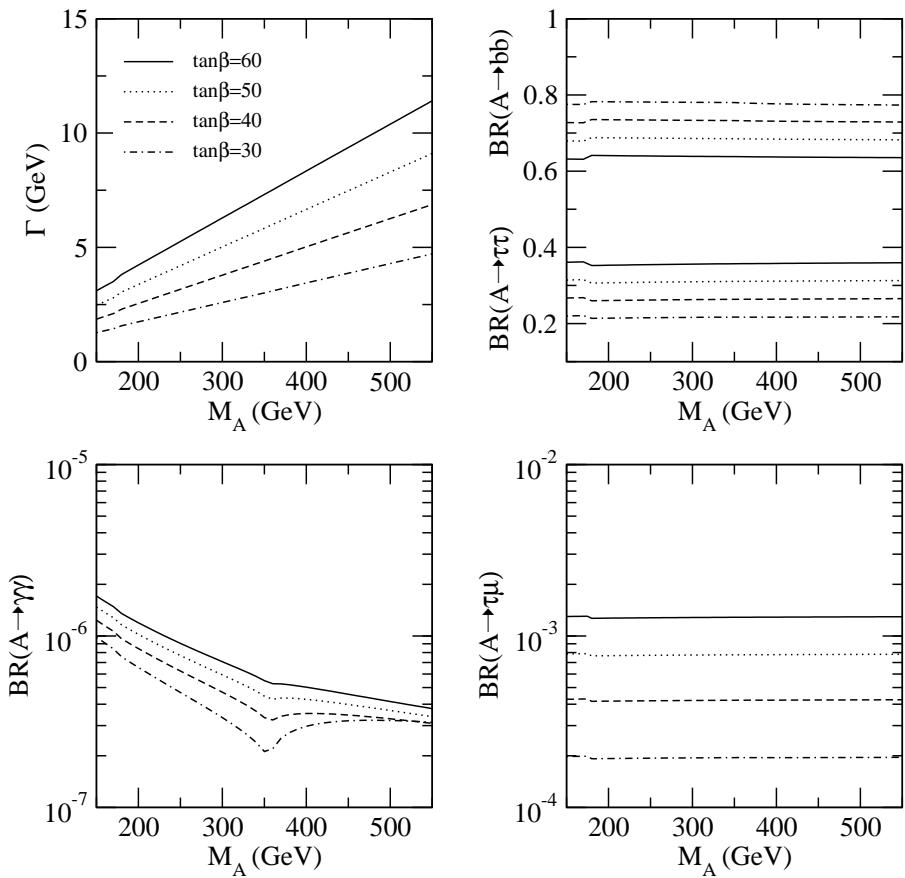}
\caption{(Top-left panel) Total width of the $A$ Higgs boson
for large values of $\tan\beta$ explained in the legend.
(Top-right panel) Branching ratios for the main decay channel
$b\bar{b}$ and $\tau^+ \tau^-$. (Bottom-left panel) Branching ratio
for $\gamma\gamma$ decay. (Bottom-right panel) Branching ratio
for the LFV decay $A \to \mu\tau$.
The results are obtained by means of the
code {\scshape{FeynHiggs}}~\cite{feynhiggs} and Eq. (\ref{hdecay})
with SUSY parameters $M_{SUSY} = M_{1,2,3} =1$ TeV, $\mu =2$ TeV, $\Delta^2 =|\Delta_L^{32}|^2 + |\Delta_R^{32}|^2 =10^{-6}$.
}
\label{higgsdecay}
\end{center}
\end{figure*}

In supersymmetry, the couplings $\Delta^{ij}$ in Eq.~(\ref{lagrangian}) are induced at one loop level
by the exchange of gauginos and sleptons, provided a source of slepton mixing is present.
In this work the analysis at Higgs LFV effects will be model independent and we use
the expressions
of $\Delta^{ij}_{L,R}$ obtained in the mass insertion (MI) approximation~\cite{paradisi1}:
\beqa
\label{deltal}
 \Delta^{ij}_{L} =
&-&\frac{\alpha_{1}}{4\pi}\mu M_1 \delta^{ij}_{LL} m_{L}^2\nonumber\\
&\times&\left[
I{'} (M_1^2, m_{R}^2, m_{L}^2)+\frac{1}{2} I{'} (M_1^2, \mu^2, m_{L}^2)
\right]\nonumber\\\nonumber\\
&+& \frac{3}{2} \frac{\alpha_{2}}{4\pi} \mu M_2 \delta^{ij}_{LL} m_{L}^2
I{'} (M_2^2, \mu^2, m_{L}^2)\ ,
\eeqa
\beq
\label{deltar}
\Delta^{ij}_{R}=
\frac{\alpha_{1}}{4\pi}\mu M_1 m^{2}_{R} \delta^{ij}_{RR}
\left[I{'}\!(M^{2}_{1},\mu^2,m^{2}_{R})\!-\!(\mu\!\leftrightarrow\! m_{L})
\right]
\eeq
where $\mu$ is the the Higgs mixing parameter, $M_{1,2}$
are the gaugino masses and $m^{2}_{L(R)}$ stands for the left-left
(right-right) slepton mass matrix entry. The LFV mass insertions $\delta^{ij}_{LL}$ and $\delta^{ij}_{RR}$ are defined as:
\begin{equation}
\label{deltas}
\delta^{ij}_{LL}\!=\!\frac{({m}^2_{L})^{ij}}{m^{2}_{L}}, \qquad \delta^{ij}_{RR}\!=\!\frac{({m}^2_{R})^{ij}}{m^{2}_{R}},
\end{equation}
where $({m}^2_{L})^{ij}$ are the off-diagonal flavor changing entries
of the slepton mass matrix. Let us emphasize that the parameters $\delta^{ij}_{LL}$ and $\delta^{ij}_{RR}$ will be treated in the following study as free parameters in order of provide a model independent study of LFV signals.
The loop function $I'$ is defined by $I'(x,y,z)= dI (x,y,z)/d z$,
where $I(x,y,z)$ is the three point one-loop integral
\begin{eqnarray}
 I (x, y, z)=
\frac{xy \log (x/y) + yz \log (y/z) + zx \log (z/x)}
{(x-y)(z-y)(z-x)}\,.
\end{eqnarray}

While gaugino mediated lepton flavor violation
decouples with the heaviest mass in the
slepton/gaugino loops $m_{SUSY}$,
Higgs mediated effects of LFV
do not decouple
increasing  the sparticles masses because $\Delta_L$, $\Delta_R$,
which appear in the couplings of the
dimension-four lagrangian (Eq.~\ref{lagrangian}), are
dimensionless coefficients given by ratios of SUSY masses.
Higgs mediated effects in rare decays
start being competitive with the gaugino mediated ones when
$m_{SUSY}$ is roughly one order of magnitude heavier then $m_H$ and
for large $\tan\beta$.
Phenomenological analysis of rare LFV $\tau$ decays and $B$ meson decays
have been widely discussed in the recent
literature~\cite{bkq,bkl,paradisi1,paradisi2,sher,dedes,brignole2,isidori1,
Parry:2005fp,
isidori2,buras,masiero3,isidori3}.

The Higgs boson decay widths and  branching ratios
relevant for the following analysis at a photon collider are
obtained by means of the lagrangian of Eq.~(\ref{lagrangian}) using the approximation $1/{c}^2_\beta
\simeq \tan^2\beta$ (only valid in the large $\tan\beta$ regime) and in the limit of  massless fermions.
Introducing  $\Delta^2 =|\Delta_L^{32}|^2 + |\Delta_R^{32}|^2$ we find:
\beq
\Gamma({A\to\tau^+\mu^-})=
\frac{1}{8\pi}\frac{m_{\tau}^2}{v^2}\, M_{A}
\, t^{4}_{\beta}\, \frac{\Delta^2}{2}.
\label{Amutau}
\eeq
where $v=(v_u^2+v_d^2)^{1/2} \approx 246$ GeV, $v_{u}$ and $v_d$ being the expectation values of the Higgs fields $H_u$ and $H_d$. The  width for the lepton flavor conserving  decay to $\tau^+\tau^-$ pair is, with the same approximations:
\beq
\Gamma({A\to \tau^+\tau^-})=
\frac{1}{8\pi}\frac{m_{\tau}^2}{v^{2}}\, M_{A} \,
t^{2}_{\beta},
\label{Atautau}
\eeq
and therefore:
\beq
\Gamma({A\to\tau^+\mu^-})=
\frac{1}{2} t^{2}_{\beta} \Delta^2  \Gamma({A\to \tau^+\tau^-}).
\label{Amutau2}
\eeq
Finally since $\Gamma({A\to \tau^+\mu^-})=\Gamma({A\to \tau^-\mu^+})$ we find also:
\beq
\Gamma({A\to \tau^+\mu^-})+\Gamma({A\to\tau^-\mu^+})
=t^{2}_{\beta} \Delta^2 \Gamma({A\to \tau^+\tau^-}),
\nonumber
\eeq
\beq
{ {\cal B}(A\to \mu^+\tau^-)}+{ {\cal B}(A\to \mu^-\tau^+)}
= t^{2}_{\beta} \Delta^2 { {\cal B}(A\to \tau^+\tau^-)}.
\label{hdecay}
\eeq

For the heavy higgs boson $H$, the right hand sides of Eq. (\ref{hdecay}),
should be multiplied by a factor $({s}_{\beta-\al}/{c}_\al)^2$. Let us remark that the light Higgs field $h$ has negligible lepton flavor violating decays since its coupling $\cos(\beta-\alpha)\to 0$ in the decoupling regime.
In Fig.~\ref{higgsdecay} we show
${\cal B}(A\to b\bar{b})$ and ${\cal B}(A\to\gaga)$ for different
values of $\tan\beta$ and for the reference point of the parameter
space $M_{SUSY} = M_{1,2,3} =1$ TeV, $\mu =2$ TeV. For completeness,
in Fig.~\ref{higgsdecay} we also report the absolute value of the
total width $\Gamma_A$ and the branching ratio for the LFV decay
$A\to\mu\tau$ given by Eq.~(\ref{hdecay}) with $\Delta^2 =10^{-6}$.\footnote{We note that our formula in Eq.~(\ref{Amutau}) agrees with the corresponding formulas in Ref.~\cite{kanemura2,Parry:2005fp}, but we find a disagreement by a factor 1/2 with that in~\cite{brignole1,moretti}.}
We note the following features: $i$) the total width is of the order of a few GeV, comparable, but always smaller, than the expected resolution of the invariant mass of the $b\bar{b}$ system (see Section~\ref{signalback}); $ii$) the total decay with is saturated almost exactly by the two decays $A\to b\bar{b}$ and $A\to \tau\tau$ while $A\to \gamma\gamma$ is strongly suppressed; $ii$) finally the branching ratio ${\cal B} (A\to \mu\tau)$ is in the interesting region of $10^{-4}\div10^{-3}$.

\section{Higgs LFV at a Photon Collider}
\label{sec3}

High-energy photons
beams~\cite{ginzburg,ILC} will be obtained from Compton
back-scattered (CB) low-energy laser photons with energy $\omega_0$
off high-energy electron beams with energy $E_e$. These high-energy
photon beams will not be monochromatic but will present instead an
energy spectrum, mainly determined by the Compton cross section. The
spectrum of the fraction of the electron's energy retained by the
photon have a maximum at  $E_{max}^{\gamma}=y_{max} E_e$, where
$y_{max} =x/(x+1)$ with $x = 4 E_e \omega_0/m^2_e$. As a
consequence, if $y_1$ and $y_2$ are the fractions of energies of the
colliding photons, the number of photon collisions as a function of
the invariant mass  $W_{\gamma \gamma}=y_1 y_2  2E_e$ will present a
spectrum too. In first approximation, the luminosity spectrum is
given by the convolution of two Compton cross-sections, that is
\beq
\frac{d L_{\gamma\gamma}^{CB}}{d z}= 2 z \,\int^{\ln{y_m
/z}}_{-\ln{y_m /z}} F_c (x,z e^{+\eta}) F_c (x,z e^{-\eta})d\eta,
\label{lumidiff}
\eeq
where we introduce the variables $z=\sqrt{y_1
y_2} = W_{\gamma \gamma} /2E_e = \sqrt{s_{\gamma \gamma}/s_{ee}}$,
$\eta = \ln{\sqrt{y_1/y_2}}$ and $F_c$ is the Compton cross section;
thus $s_{\gaga} =z^2 s_{ee}$ with $s_{ee}=(2E_e)^2$.
Even if the simulated realistic differential spectrum as a function
of $z$ cannot be described analytically, the
luminosity peak near $z_{max}$ is almost independent from the
details of the machine, and is well described by
Eq.~(\ref{lumidiff}). We consider a PC taking the parameters of
TESLA(800): $2E_e =800$ GeV, $x=5.2$. $2E_\gamma$ in the region of
the peak covers the range $535-670$ GeV, and the corresponding
photon-photon luminosity at the peak is $L_{\gamma\gamma}(z>0.8z_m)=
1.7 \times 10^{34}$ cm$^{-2}$ s$^{-1}\simeq 500$ fb$^{-1}$ yr$^{-1}$.
To use these numbers as representative
of the PC luminosity, we follow the approach of Ref.~\cite{Cannoni:2005gy}, in
which the ideal spectrum, Eq.~(\ref{lumidiff}), is normalized in the
following way:
\beqa
\frac{dL^{norm}_{\gamma\gamma}}{dz}&=&
\frac{1}{\int^{z_{max}}_{0.8z_{max}}dz \frac{dL^{CB}_{\gamma\gamma}}{dz}}
\frac{dL^{CB}_{\gamma\gamma}}{dz},
\eeqa
Defining the effective cross section as:
\beq
\sigma^{eff}=\int^{z_{max}}_{z_{min}}dz
\frac{dL^{norm}_{\gamma\gamma}}{dz} \sigma(W_{\gamma\gamma}),
\label{eff}
\eeq
the total number of events can be evaluated to be
$N_{events}=L_{\gamma\gamma}\times\sigma^{eff}$, where
$L_{\gamma\gamma}$ is the simulated TESLA luminosity at the peak.

In photon-photon collisions the main production mechanisms for the Higgs bosons
are $\gamma\gamma$ fusion~\cite{djouadi1} and $\tau\tau$ fusion~\cite{choi}.
In the first the Higgs is produced as an s-channel resonance through a loop involving the exchange of all charged massive particles.
This process is particularly well suited for precision studies of the Higgs sector when the mass(es) of the Higgs boson(s) is (are) known,
given that the width of the decay into two photons (which controls the production cross section)
is not too small.
Assuming that the machine  energy could be tuned to the Higgs mass then one could take
advantage of the resonant production in order to probe the details of the  Higgs sector.
However in the case of $\tan\beta$ enhanced Higgs-lepton coupling the $\tau\tau$ fusion
becomes competitive or even dominant over a wider range of masses.
In order to appreciate the differences between the two mechanisms, we first
give analytical expressions of the cross sections for the LFV Higgs
mediated signals and then make numerical estimates. For clarity, we fix the
following SUSY parameters to: $M_{SUSY} = M_{1,2,3} =1$ TeV, $\mu
=2$ TeV, $\tan\beta =50$, $\Delta^2=|\Delta_L^{32} |^2 +|\Delta_R^{32} |^2 = 10^{-6}$,
as for the calculation reported in Figure~\ref{higgsdecay}; the widths and branching ratios which
appear in the following formulas are computed with the software
{\scshape{FeynHiggs}}~\cite{feynhiggs} linked to our own code.
\begin{figure*}[t!]
\begin{center}
\includegraphics*[scale=0.65]{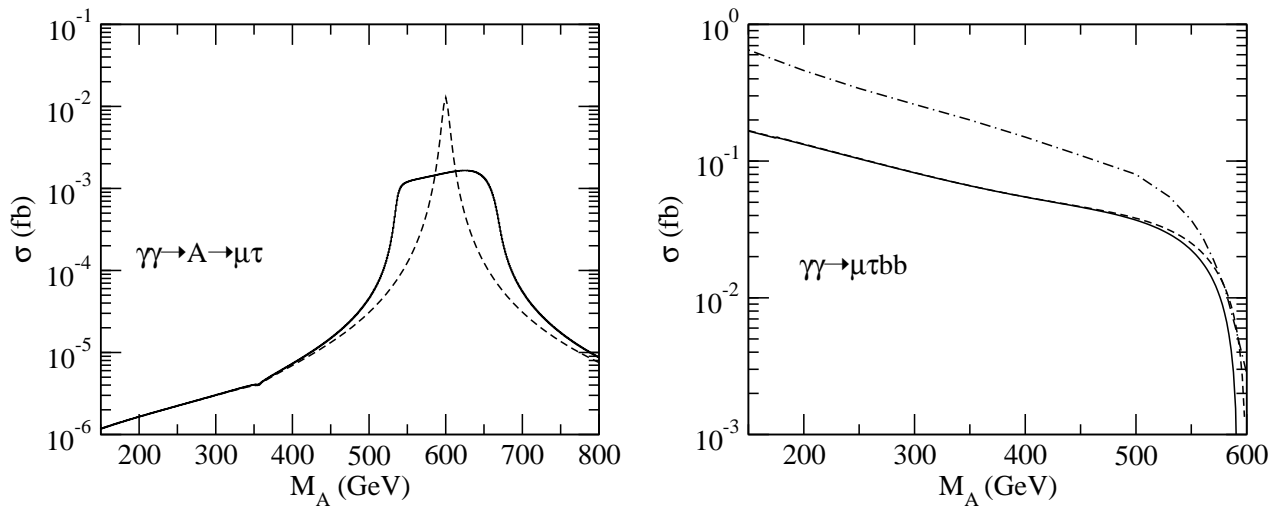}
\caption{ (Left panel) Cross section for $\gamma \gamma \to
A \to \mu\tau$ as a function of $M_A$: dashed line
monochromatic photons, in full line the effective cross
section with photons luminosity. (Right panel) Cross
section for $\gamma\gamma \to \mu\tau b\bar{b}$ . Full
line: monochromatic photons and equivalent particle
approximation plus small width approximation, Eq.
(\ref{fusionfinal}). Dashed line: effective cross section,
Eq. (\ref{fusionfinalspectra}). Dot-dashed line: exact
result by \textsc{CompHeP}. The relevant parameters are
$\tan\beta =50$, $|\Delta_L^{32} |^2 +|\Delta_R^{32} |^2 = 10^{-6}$,
$M_{SUSY} = M_{1,2,3} =1$ TeV, $\mu =2$ TeV.}
\label{analitica}
\end{center}
\end{figure*}

The cross section for the resonant process $\gamma \gamma \to A \to
\mu\tau$ in the monochromatic case is provided by a Breit-Wigner formula~\cite{djouadi1}:
\begin{equation}
{\sigma}(s_{\gaga}) =  8\pi \, \frac{\Gamma(A\to \gamma \gamma) \,
\Gamma(A \to \tau\mu)}{(s_{\gaga}-M_A^2)^2 + (\Gamma_{A}M_A)^2} \,
(1+\lambda_1\lambda_2) \label{sigmaannhi}
\end{equation}
where $\lambda_{1,2}$ are the photons helicities and we take
$\lambda_1\lambda_2 =1$. The effective cross section is obtained
by folding Eq.(~\ref{sigmaannhi}) in Eq.~(\ref{eff}). In Fig.~\ref{analitica} we plot
with a dashed line the monochromatic cross section with $2E_\gamma
=600 $ GeV and with a  full line the effective cross section with $2E_{e}
=800$ GeV: the resonance peak at $M_A =600$ GeV is only smoothed by
the effect of the photon spectra, while it is clear that in mass regions away from the
resonance the effect of the photon spectra is totally negligible. Even if the resonant effect is evident
around $M_A =600$ GeV, the cross section is rather small, at the level
$10^{-2}\div10^{-3}$ fb, being suppressed by $\Gamma(A\to
\gamma\gamma)$ which in our scenario is $\cal{O}$($10^{-6}$) GeV, as can be
seen in Fig.~\ref{higgsdecay}.
\begin{figure*}[ht!]
\begin{center}
\includegraphics*[scale=0.65]{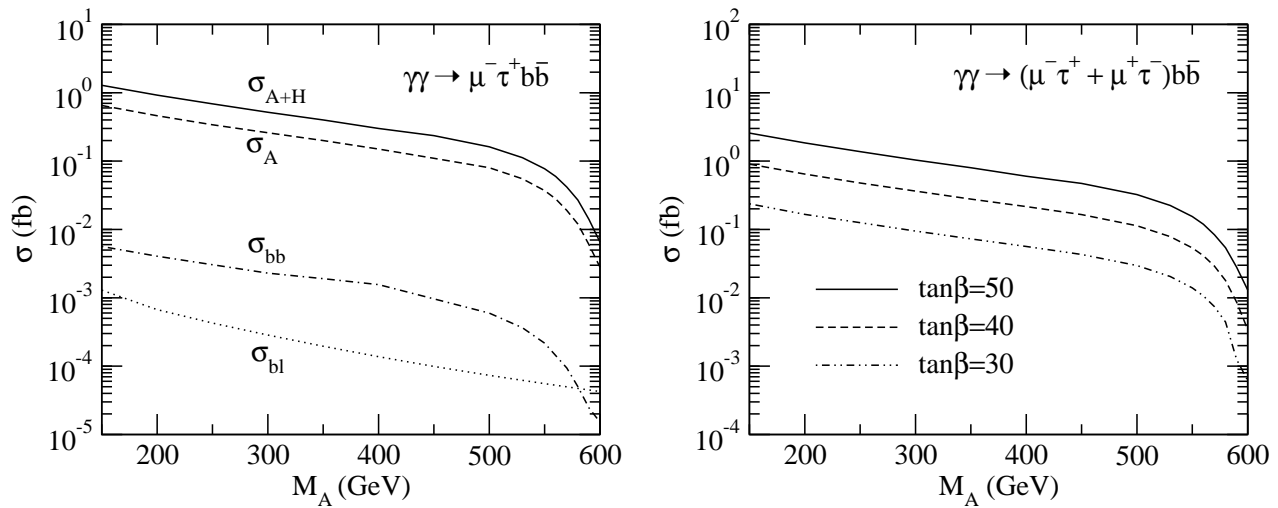}
\caption{(Left-panel) Total cross for $\gamma\gamma \to
\mu\tau b\bar{b}$ with monochromatic photons at $2E_\gamma
= 600$ GeV as a function of $M_A$ and $\tan\beta =50$. Full
line: all contributing diagrams with $A$ and $H$. Dashed
line: only diagrams with $A$. Dot-dashed line: $b\bar{b}$
fusion cross section. Dotted line: contribution of the
peripheral diagrams. (Right panel) Exact cross section
with $A$ and $H$ diagrams for different values of the
$\tan\beta$. The other parameters are the same as in Fig.
(\ref{analitica}).  } \label{sigtot}
\end{center}
\end{figure*}

The $\mu\tau$ fusion process $\gamma\gamma \to \mu\tau b\bar{b}$
corresponds to the  diagram $(a)$ in Fig.~\ref{diagrams}. Here
the Higgs boson is produced in the s-channel via a virtual $\mu\tau$ pair and
can be detected from its decay mode $A \to b\bar{b}$. The black blob in
the vertex of the diagram represents the loop induced LFV coupling.
A first estimate of the production cross section can be obtained using
the equivalent particle approximation (EPA) wherein the colliding
real photons split into $\tau$ and $\mu$ pairs with the subsequent
$\mu\tau$ fusion into the Higgs boson.
Following Ref.~\cite{choi}, we introduce the photon splitting function into a pair of leptons
\beq
P_{\gamma/\ell}(x)=\frac{\alpha}{2\pi}[x^2 +(1-x)^2]
\ln\left(\frac{\mu_F^2}{m_\ell^2}\right)
\label{splittingfunction}
\eeq
where $x$ is the fraction of the energy of the photon carried by the
virtual lepton and $\mu_F $ is the factorization scale that we set to $\mu_F=M_A$.
The on shell $\mu\tau \to b\bar{b}$ fusion cross section in the center of mass frame
is easily calculated from the lagrangian of Eq.~(\ref{lagrangian})
and expressed in terms of the partial  widths given in Section~\ref{sec2}:
\beq
{\sigma}(\hat{s})= \frac{ 4\pi \Gamma({A\to\tau\mu})
\Gamma (A\to b\bar{b}) \hat{s}} { M_A^2
(\hat{s}-M_A^2)^2 + (\Gamma_{A} M_A)^2}.
\label{mutaufusion}
\eeq
where $\hat{s}$ is $\mu\tau$ center of mass energy squared.
The cross section for monochromatic photons is given by the
convolution of Eq.~(\ref{mutaufusion}) with the splitting functions,
\beq
\sigma(\gaga \to \mu\tau b\bar{b}; s_{\gaga})= 2\int dx_{\mu}
dx_{\tau} P_{\gamma/ \mu}(x_\mu)  P_{\gamma/\tau}(x_\tau)
{\sigma}(\hat{s}),\nonumber
 \label{gagamutauA}
\eeq where $s_{\gaga}$ is the center of mass energy squared
of the photons which is related to $\hat{s}$ by
$\hat{s}=x_{\mu} x_{\tau} s_{\gaga}$ and the factor two is
the multiplicity factor which accounts for the exchange of
the initial photons. The formula is simplified by changing
the  integration variables $(x_\mu,x_\tau)$ to $(\eta,
\hat{s})$ with $\eta = \ln\sqrt{{x_\mu}/{x_\tau}}$  and
using the small width approximation (SWA) when performing
the integration over $\hat{s}$. The result is: \beqa
&&\sigma(\gaga \to \mu\tau b\bar{b}; s_{\gaga})= \frac{4
\pi^2}{s_{\gaga}}
\frac{\Gamma(A\to \tau\mu) {\cal B}(A\to b\bar{b})}{M_A}\cr
&&\times 2\int_{-\ln 1/t}^{+\ln 1/t} {d\eta}\; P_{\gamma/ \mu}\left(t e^\eta \right)
P_{\gamma /\tau}\left(t e^{-\eta}\right).
 \label{fusionfinal}
 \eeqa
with $t={M_A}/{2E_\gamma} $. Finally, the effective cross section is
obtained by the convolution of Eq. (\ref{fusionfinal}) with the
photon spectra; defining $t ={M_A}/{2E_e}$ we have
\beqa
\sigma_{eff}&=&
\frac{4 \pi^2}{s_{ee}}
\frac{\Gamma(A\to \tau\mu) {\cal B}(A\to b\bar{b})}{M_A}\cr\cr &\times&
\left[ \int_{z_{min}}^{z_{max}}dz\,
\frac{dL_{\gamma\gamma}^{norm}}{dz}\right. \cr
&\times& \left. 2\int_{-\ln 1/zt'}^{+\ln 1/zt'} {d\eta} \; P_{\gamma /\mu}\left(\frac{t'}{z}
e^\eta \right) P_{\gamma /\tau}\left(\frac{t'}{z}e^{-\eta}\right)
\right] \nonumber \\
\label{fusionfinalspectra}
\eeqa
In Fig.~\ref{analitica} we plot the Eq.~(\ref{fusionfinal}) and
Eq.~(\ref{fusionfinalspectra})
as functions of $M_A$. The cross section is in the range $10^{-2} - 1$
fb in the range of Higgs masses $100 - 550$ GeV, thus dominant
respect to the $\gamma\gamma$ fusion one. Moreover, the use of the
photon luminosity spectrum with $2E_e = 800$ GeV and $x=5.2$ gives
the same numerical results for the cross section calculated with
monochromatic photons with $2E_{\gamma}=600$ GeV which represents
the value of the mean energy at the luminosity peak, $535$ GeV $\leq
2E_{\gamma}\leq 670$ GeV, so that from now on we consider this
situation of monochromatic photon beams to simplify the
calculations.

To estimate the accuracy of the analytical formulas we also show in
Figure~\ref{analitica} (right plot) the cross section calculated
with Eq.~(\ref{gagamutauA}) and the one calculated with the program
{\scshape{Comphep}}~\cite{comphep} in which we have implemented the
MSSM with LFV as described by the lagrangian in
Eq.~(\ref{lagrangian}).
The dotted dashed curve is obtained considering
the $\mu\tau$ fusion diagrams  and
bremsstrahlung diagrams with the $A$ boson contribution,
Figure~\ref{diagrams}(a-b),
which, as we show explicitly in Section \ref{signalback},
are the dominant diagrams.
The analytical result gives the correct order of magnitude of the
cross section, but for low Higgs masses it underestimates the exact
result by a a factor 3--5, and only for masses approaching to
$2E_\gamma$, the kinematical limit, the approximation is better.
This analytical study served us to provide a preliminary evaluation of the
orders of magnitude of the signal and to understand  the dominant
mechanisms involved. In the following Section~\ref{signalback}
we present the results of full numerical  tree-level simulations
obtained  by means of {\scshape{Comphep}}.
\begin{figure*}[t!]
\begin{center}
\includegraphics[scale=0.7]{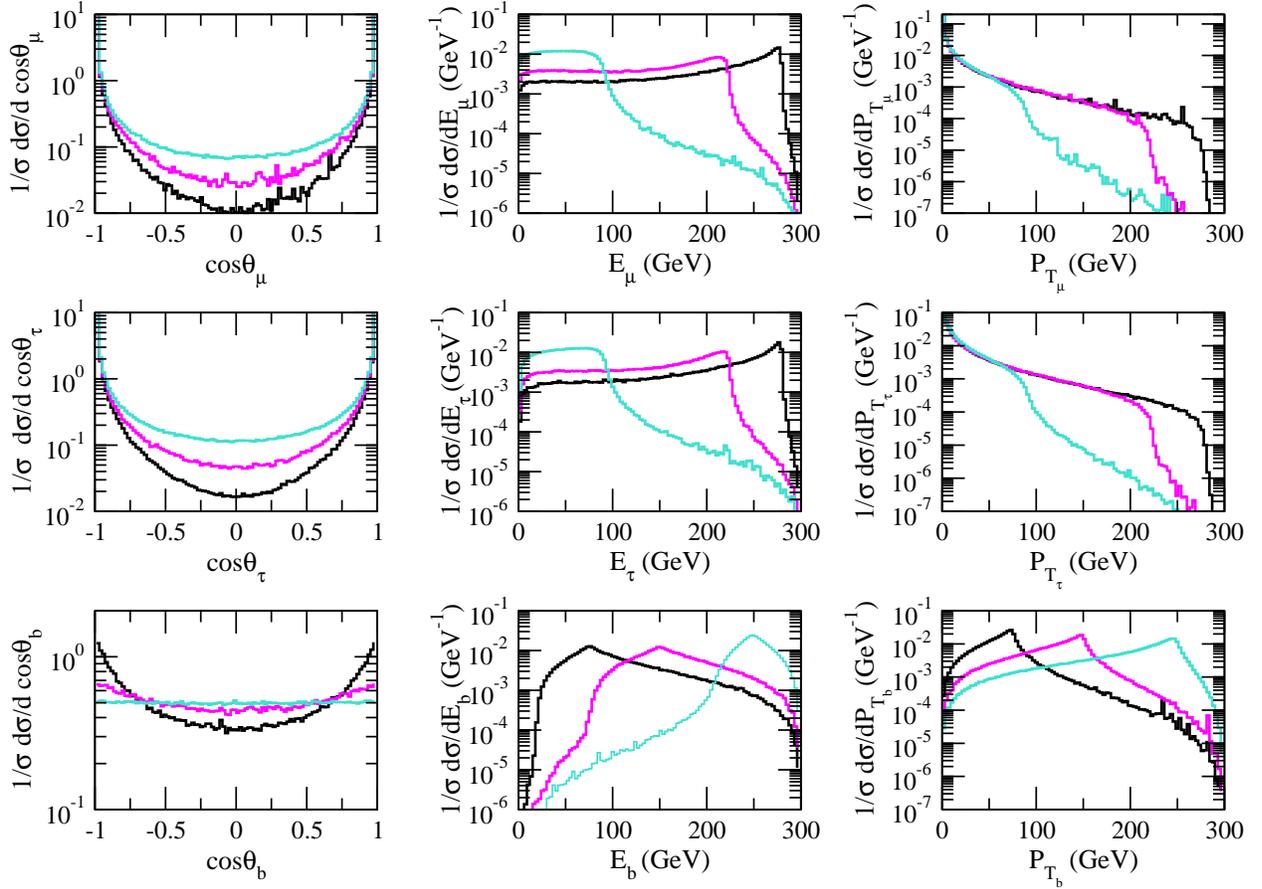}
\caption{(Left column) Distributions for scattering angle
respect the collision axis for leptons and jets. (Central
column) Distributions for the energy of leptons and jets.
(Right column) Distributions for the transverse momentum of
leptons and jets. Black line: $M_{A,H} = 150$ GeV, Magenta:
$M_{A,H} = 300$ GeV, Cyan: $M_{A,H} = 500$ GeV. The other
parameters are the same of Figs. (\ref{analitica}). }
\label{sig1}
\end{center}
\end{figure*}

\section{Signal and background}
\label{signalback}
The process $\gaga \to \mu\tau b\bar{b}$
mediated by the heavy Higgs bosons $A$ and $H$ is described
by a set of 40 diagrams which can be classified by the
three topologies depicted in Figure~\ref{diagrams}: (a) the
$\mu\tau$ fusion diagrams where the Higgs is in the $s$
channel; (b) bremsstrahlung diagrams, where the Higgs is
radiated by a lepton of an external leg; (c) peripheral
diagrams, where the Higgs bosons are exchanged in the
$t-(u)$ channel. In our numerical calculations we have
divided  them into three groups: (group-1) 12 diagrams
(topology (a) and (b)) describe what we call ``$\mu\tau$
fusion'' to $A,H$; (group-2) 12 diagrams which
 describe ``$b\bar{b}$ fusion'' to
$A,H$ (the are given by the topologies (a) and (b) of
Fig.~\ref{diagrams} with the role of $\mu\tau$ and
$b\bar{b}$ excahnged); (group-3) 26 diagrams of topology
(c) that we call ``$b l$ fusion''.

In Figure~\ref{sigtot}, left panel, we plot the
contribution to the total cross section of these groups,
$\sigma_{\mu\tau}$, $\sigma_{bb}$ and $\sigma_{bl}$,
respectively. We observe the following features:
$\sigma_{bb}$, even if it is described by diagrams with the
same phase-space structure of $\sigma_{\mu\tau}$, is two
orders of magnitude smaller because those diagrams with two
$b$-quark attached to photon lines bring in a charge factor
of $(1/3)^2$ in the amplitude; $\sigma_{bl}$ is three
orders of magnitude smaller of $\sigma_{\mu\tau}$, both for
the presence in the diagrams of at least one
$b\bar{b}\gamma$ coupling and the absence of $s$-channel
resonant propagators; finally we note that in
$\sigma_{\mu\tau}$ the contributions of diagrams with $A$
and  $H$ sum up incoherently, i.e. their interference
vanishes. Since in the limit of large $\tan\beta$,
$M_A \approx M_H$ and also the couplings of the higgs bosons
$A$ and $H$ become approximatively equal  we have
$\sigma_{A+H}\approx 2\sigma_{A}$.
We conclude that the signal cross section is completely
determined by $\sigma_{\mu\tau}$, while $\sigma_{bb}$ and
$\sigma_{bl}$, which are irreducible backgrounds are
negligible. (We also checked that  the interference of
these subleading contributions with the dominant one is
negligible.)

\begin{figure*}[t!]
\begin{center}
\includegraphics[scale=0.75]{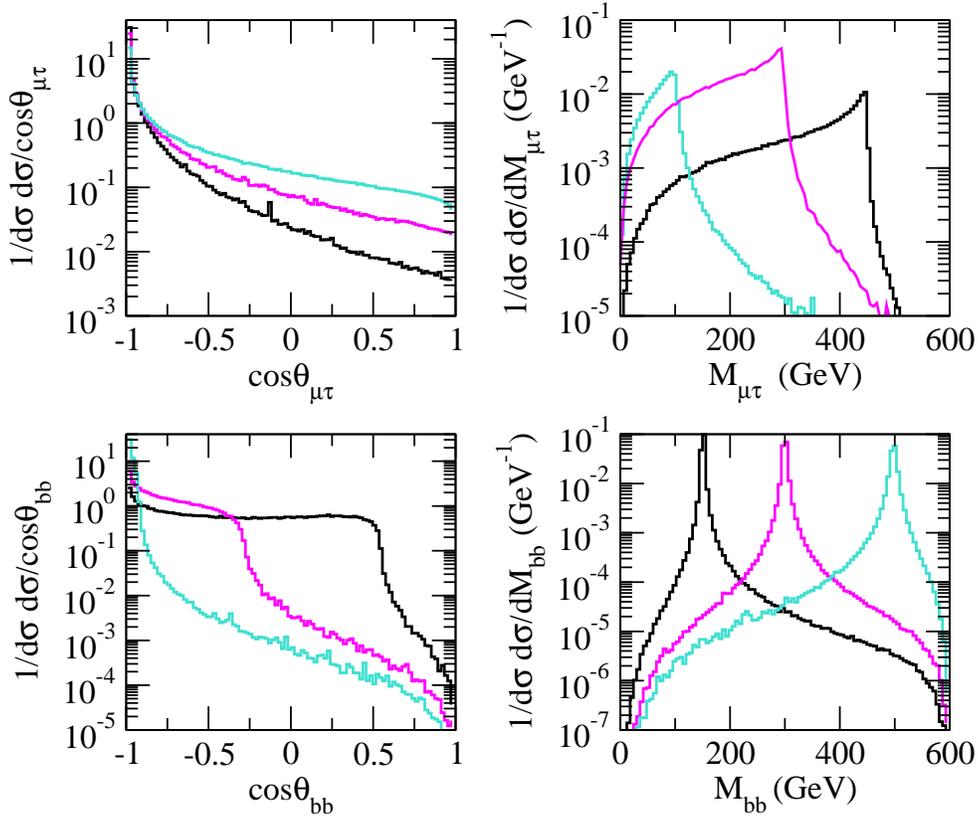}
\caption{ (Left column) Distributions for opening angle
between the two leptons and the jets. (Right column)
Distributions for the invariant mass of leptons pair and
$b\bar{b}$. Black line: $M_{A,H} = 150$ GeV, Magenta:
$M_{A,H} = 300$ GeV, Cyan: $M_{A,H} = 500$ GeV. The other
parameters are the same of Fig. (\ref{analitica}).}
\label{sig2}
\end{center}
\end{figure*}
The right panel of Figure~\ref{sigtot}  shows the total
cross section $\sigma_{\mu\tau}$ for $\tan{\beta}$ in the
range $ 30< \tan\beta < 50$ as a function of $M_A$; a
factor of two as been included to account for the charged
conjugated process which has the same total cross section.
With $\Delta^2= |\Delta_L |^2 +|\Delta_R |^2 = 10^{-6}$ the
cross sections range from $10^{-2}$ fb to $2$ fb with $M_A$
ranging from $150$ GeV to $\approx 550$ GeV. Assuming an
integrated luminosity for the photon collider in the range
$200-500$ fb$^{-1}$/yr, a relatively large number of events
are thus predicted, from ${\cal N}_{events} \approx 200
-500$ for $M_A = 100$ GeV down to  ${\cal N}_{events}
\approx 20 -50$ when $M_A \approx 550$ GeV. However, we
show in the next section, that these numbers are rather
optimistic. The results on the collider cross sections must
be correlated with  the constraints on the Susy spectrum,
$B$-physics and LFV $\tau$ decays are considered. Once this
is done the allowed parameter space is reduced and more
realistic predictions emerge.

In Fig.~\ref{sig1}, left column, are  shown the angular
distributions as function of  the cosine of the angle
between one particle ($\mu, \tau,  jet$) with the  positive
direction of the collision axis for three values of
$M_{A,H}=(150, 300,500)$ GeV. The distribution is peaked
along the collision axis for the leptons and the effect is
more pronounced at low Higgs masses, while for jets the
cross cross section is less concentrated in the
forward-backward directions and is practically flat for
$M_{A,H}$ above $300$ GeV. The distributions in the
transverse momentum (right column) are consistent with the
angular distributions, leptons have low $p_T$, the quarks
have high transverse momentum with distribution  peaked
around $M_{A,H}/2$. For the b-jets also the energy
distribution (central column) is peaked at $M_{A,H} /2$,
while leptons are more energetic for low Higgs masses.
Other interesting features of the signal are given by the
distribution for the cosine of the angle between the
leptons and jets shown in Fig.~\ref{sig2}, left column.
Both are peaked towards $\cos\theta_{ij}\to -1$, thus both
the lepton pair and the jet pair will be well separated
being almost back-to-back, the effect being stronger for
the b-tagged jets at high $M_{A,H}$. On the right column,
we plot the invariant mass distribution for $\mu\tau$ pair
and the $bb$ pair: while the former extends up to
$\sqrt{s_{\gamma\gamma}}- M_{A,H}$, the latter has a peak
at the Higgs mass, as expected, because of the s-channel
propagator in the amplitude. Thus the signal has the
following characteristics: the Higgs decay to a pair of
back to back $b$-jets with energy and transverse momentum
around $M_A /2$ and invariant mass peaked at $M_A$. The
$\mu$ and the $\tau$ are also back to back and in
forward-backward directions with low $P_T$.
\begin{table*}[ht!]
\caption{Effect of cuts on the cross section for
$\gamma\gamma \to \mu \tau b\bar{b}$: $\sigma^{cut}$ is
obtained imposing for all  particles in the final state: scattering angle
$\theta >130$ {mrad}, $E>5$ GeV, $b\bar{b}$ invariant mass
$M_A -5\% M_A < M_{b\bar{b}}< M_A +5\% M_A $. In
$\sigma^{cut}_{P_T}$ a cut on transverse momentum of
leptons $p_T >5$ GeV is added.}
\begin{ruledtabular}
\begin{tabular}{lcccc}
$M_{A}\,$ (GeV)\phantom{\LARGE $X_{P_T}$} & $\sigma^{cut} \, $ \ (fb) &
$\sigma^{cut}_{P_T} \, $ \ (fb)&$\sigma^{cut} \, $ \ (fb)
&$\sigma^{cut}_{P_T} \, $ \ (fb)\cr\hline \phantom{\LARGE $X_{P_T}$}&
$\tan\beta =40$&$\tan\beta =40$
&$\tan\beta=50$& $\tan\beta=50$\cr\hline
150&0.240&0.124&0.660&0.340\cr 200&0.186&0.096&0.520&0.280\cr
300&0.122&0.074&0.340&0.172\cr400&0.070&0.042&0.240&0.160\cr
500&0.052&0.024&0.148&0.076\cr
\end{tabular}
\end{ruledtabular}
\end{table*}

The background processes from the SM are the ones with the final
state $\mu\tau b\bar{b}$+neutrinos. The main processes are:
\begin{eqnarray}
&&(a)\,  \gaga \to Z^* Z^*\to (b\bar{b})(\tau^+ \tau^{*-})\to
b\bar{b}\tau^+ \mu^- \bar{\nu_\mu}\nu_\tau  \nonumber\\
&&(b)\,  \gaga \to W^{*+} W^{*-} Z^* (\gamma^*)
\to (\tau^+ \nu_\tau)(\mu^- \bar{\nu_\mu})(b\bar{b})\,.\nonumber
\end{eqnarray}
The cross sections for double and triple gauge boson
production are known to be large in photon-photon
collisions~\cite{ILC}. At $\sqrt{s_{\gaga}}=600$ GeV they
are: $\sigma(\gaga \to ZZ)=0.2$ pb and $\sigma(\gaga \to
WWZ)=0.7$ pb.  We estimate the cross section for the
complete processes by multiplying the above numbers by the
appropriate branching ratios for the decay chains and find:
$\sigma_a \simeq 1.77\times 10^{-1}$ fb and $\sigma_b
\simeq 1.26$ fb. The jets coming from $Z$ decay have very
different distributions and energies from those of  the
jets from Higgs decay. Moreover, while the signal has no
missing energy or missing transverse momentum, the two
neutrinos in the final state of the SM backgrounds carry
away a large fraction of the energy/momentum, thus
providing large missing energy and momentum.

As the bulk of the cross section is determined by those
regions of phase space where the leptons are emitted with
small angle and low transverse momenta and therefore might
escape detection, it is essential to evaluate the expected
number of events including  angular cuts. We apply a cut of
$\theta >130$ {mrad} on the scattering  angles of leptons
and jets for detector acceptance and further a cut on the
energy of both leptons and jets: $E>5$ GeV; the invariant
mass of the $b\bar{b}$ system is required to be in the
range $M_A \pm 0.05M_A$, which is the expected experimental
resolution~\cite{choi}. We find that after the cuts the
background processes have cross sections at the level of
$10^{-3}-10^{-4}$ fb, while the effect on the signal cross
section can be read off from Table I. In particular we
observe that the cut on the  invariant mass $M_{bb}$
suffices to suppress background processes.
Comparing the numerical results shown in Table I
($\sigma^{cut}$) with Fig.~\ref{sigtot} we note that the
signal cross section is reduced by a factor of three;
moreover we show also the effect of a supplementary cut on
transverse momentum of leptons $p_T > 5$ GeV which reduces
the cross section ($\sigma^{cut}_{P_T}$) by another factor
of two as one might expect given the transverse momentum
distributions. On the contrary, we find that the cut on the
energy can be raised up to 50 GeV lowering the cross
section only of $\approx 20\%$.

Another source of background are fake events where the muon
comes from the decay of a $\tau$ in the lepton flavor
conserving process $\gamma\gamma \to \tau^+ \tau^-
b\bar{b}$: this process has the same characteristic
of the signal, and will pass the above cuts.
The used version of {\scshape{Comphep}}
allows the study of six particle final states,
and works properly if we restrict the numbers of contributing
diagram: thus we considered the dominant diagrams in $\tau\tau$
fusion with one $\tau$ decaying $\tau \to \nu \bar{\nu} \mu$.
We find, without any cut, cross sections
$4\times 10^{-2},\,2.4\times 10^{-2},6.2\times 10^{-3},\,
7.2\times 10^{-4},\,2.1\times 10^{-5}$ fb
for masses $M_{A,H}=150,\,200,\,300,\,400,\,500$ GeV respectively.
The suppression and the rapid fall with increasing $M_A$ can be
understood kinematically: the tree-body decay of the very
energetic $\tau$ is disfavored for phase space reasons,
especially at large $M_A$ where the $b$ are more energetic,
as discussed above, and less energy is sheared by the leptons.

We finally remark that the LFV channel we are considering
would naturally be a second stage study after a thorough investigation of  the  LFC $\tau\tau$ channel is performed,
which should  allow precision studies of the Higgs sector ($M_{H,A},\tan\beta$ etc.).
Thus, once the mass and other properties are known, the search
for the LFV channel would be facilitated: for example, the fact that the invariant
mass of the leptons is peaked at $2E_{\gamma}-M_A$ may be useful
in selecting the signal (though the center of mass energy of the
PC is not exactly fixed, as discussed above, at the luminosity
peak the photons are almost monochromatic).

\section{Correlations with  low energy constraints }
\label{sec5}
\begin{figure*}[t!]
\begin{center}
\includegraphics[scale=0.7,angle=-90]{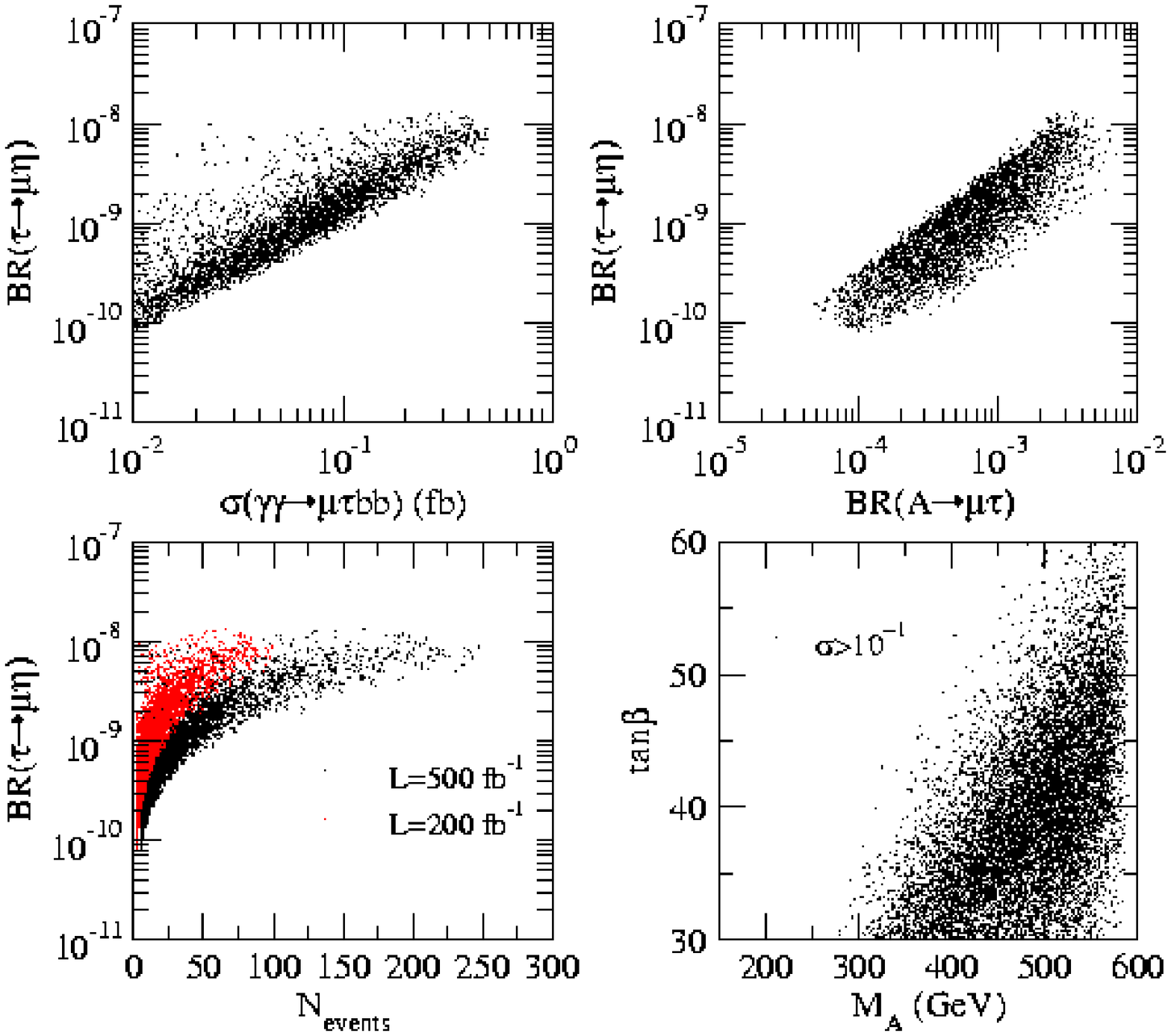}
\caption{ (Top left panel) Correlation between ${\cal
B}(\tau\to\mu\eta)$ and $\sigma(\gaga \to \mu\tau bb)$.
(Top right panel) Correlation between  ${\cal
B}(\tau\to\mu\eta)$ and $A \to \mu \tau$. (Bottom left
panel) Correlation between ${\cal B}(\tau\to\mu\eta)$ and
the number of expected events for two values of the
integrated luminosity. (Bottom right panel) Distribution of
the signal cross section in the ($M_A , \tan\beta$) plane.
The parameter space and the imposed constraints are discussed in the text.
} \label{lowvshigh}
\end{center}
\end{figure*}

The Higgs bosons LFV vertices and the
branching ratios depend on the parameters of the theory
(MSSM+LFV) that are subject to non-trivial constraints by
experiments. In order to provide a detailed study of the
possibilities of a photon collider with respect to the LFV
violating signal $\gamma \gamma \to \mu\tau b \bar{b}$ we
scan over the following parameter space:
\begin{eqnarray}1\, \hbox{TeV} &\leq&(\mu,
m_{\tilde{q}}, A_u, A_d, m_{{L}},m_{{R}})\leq5\, \hbox{TeV},\cr 500\,  \hbox{GeV} &\leq&(M_1 , M_2, M_3)\leq 5\, \hbox{TeV},\cr
150\, \hbox{GeV} &\leq& M_A \leq1\, \hbox{TeV},\cr 30\,&\leq& \tan\beta \leq\, 60,\cr
10^{-3}&\leq&\, (\delta_{LL}^{32},\delta_{RR}^{32})
\leq0.5.
\end{eqnarray}
The parameters $(\delta_{LL}^{32},\delta_{RR}^{32})$, defined in Eq.~\ref{deltas},
measure, in a model independent way,  the amount  of lepton flavor violation.
We verified that
$\delta_{LL}^{32}$, which give the most important contribution,
should be greater than $\simeq 5\times
10^{-2}$ to have substantial cross section.

We impose the SUSY
parameter space to respect the following constraints: lower
bound on the light Higgs mass $m_h >114.4$ GeV; upper bound
on the anomaly of the muon magnetic moment  $(g-2)_{\mu}< 5\times 10^{-9}~$;
bounds on electro-weak
precision observables such as $\Delta\rho<1.5\times
10^{-3}$; direct search constraints on the lightest
chargino and sfermion masses and constrains on squarks and
gluino masses from LEP and Tevatron are automatically
satisfied as they lie in the TeV range in our scenario.

Some $B$-physics processes, namely $B_s \to \mu^+ \mu^-$,
$B \to X_s \gamma$ and $B_u \to \tau \nu$, are particularly
sensitive to $\tan\beta$ enhanced Higgs contributions which
have been subject of extensive recent
studies~\cite{isidori1,isidori2,buras,isidori3}. In
particular: using the formulas for the branching ratio
given in Ref.~\cite{buras} we require that the parameter
space satisfies ${\cal B}(B_s \to \mu^+ \mu^-)<6.5\times 10^{-8}$
\cite{pdg}; $R_{\tau\nu}$, the ratio between the SUSY and
SM branching ratios for $B_u \to \tau \nu$, is required in
the bracket $0.70<R_{\tau\nu}<1.44$ using the formula in
Ref.~\cite{isidori3} and the numerical bounds
from~\cite{masiero3}; $R_{X_s\gamma}$, the ratio between
the SUSY and SM branching ratios for $B \to X_s \gamma$, is
required to lie in the bracket $1.01<R_{X_s\gamma}<1.25$
through the formulas of Ref.~\cite{isidori1} and the
numerical bounds are taken from~\cite{masiero3}.

At last, we impose the current  upper bounds on LFV $\tau$
decays to be respected: ${\cal B}(\tau^- \to \mu^-
\eta)<6.8\times 10^{-8}$ and ${\cal B}(\tau^- \to \mu^-
\gamma)<5.6\times 10^{-8}$~\cite{pdg}. In the case where
Higgs-mediated LFV effects are important,
$\tau\to\mu\eta$ is generally the dominant
process~\cite{sher,brignole2,paradisi2}.

We employ the
analytical formulas of Section II for the cross section
which gives a reasonable estimate when cuts are taken into account.

In Fig.~\ref{lowvshigh}, top-left panel, we show the
correlation between ${\cal B}(\tau\to\mu\eta)$ and
$\sigma(\gaga \to\mu\tau b\bar{b})$, while in the
bottom-left panel, the correlation between ${\cal
B}(\tau\to\mu\eta)$ and the numbers of $\mu\tau b\bar{b}$
events given by the previous cross section available at a
photon collider for two values of the integrated
luminosity, ${\cal L}=200-500$ fb$^{-1}$/yr. It can be seen
that for the high luminosity option we can expect up to 250
events per year, and up to 100/yr for the low luminosity
option. The above conclusions are valid for the present
upper limits on the branching ratios: if in the near future
the experimental upper bound will be improved (i.e.
lowered), say by an order of magnitude, only few tens of
events can be expected unless higher values of luminosity
should become in the meantime available.

In the bottom-right panel we show the region of the parameter space in the $(M_A ,\tan\beta)$ plane  which is characterized by a signal cross-section $\sigma \geq 10^{-1}$ fb.
Let us remark that  the signal cross-sections becomes larger at low $M_A$ masses, see Fig.~\ref{sigtot} and Eq.~(\ref{gagamutauA}). However in the considered region of large $\tan\beta$ values such low masses are excluded by the imposed constraints.
In particular,
the LFV signal for $M_A$ masses below $300$ GeV are
excluded for all values of  $\tan\beta$ in the interval,
$30 <\tan\beta<60$.
We have checked that requiring a signal cross section $10^{-2} \text{fb} \leq \sigma \leq 10^{-1}$ fb
the same region in the $(M_A ,\tan\beta)$ plane is covered.

Finally, in the top-right panel we show the correlation
between ${\cal B}(\tau\to\mu\eta)$ and ${\cal
B}(A\to\mu\tau)$. The latter gets values in the interval $(5\times10^{-4})\lesssim {\cal
B}(A\to\mu\tau) \lesssim (8\times 10^{-3})$. Even if
the upper limit on  ${\cal B}(\tau\to\mu\eta)$ is lowered by an order of
magnitude form its actual value ($\approx 10^{-8}$) we see that ${\cal
B}(A\to\mu\tau)$  can still reach values up to ${\cal O}(10^{-3})$ which are
particularly interesting for the  LHC, where
the cross section for heavy neutral gauge bosons production
in $b\bar{b}$ fusion is sizable~\cite{moretti}.

\section{Summary and conclusions}
\label{sec6}

In this work we present a detailed study of
LFV signals at a photon collider, namely $\gamma \gamma \to
\mu \tau b \bar{b}$, mediated by the MSSM heavy Higgs
bosons ($A,H$).  Our approach is model-independent with
respect to the source of lepton flavour violation.
Effective couplings of the MSSM higgs bosons which violate
lepton flavor conservation  arise at loop level once in the
model (MSSM) it is introduced a source of LFV in the
slepton mass matrix. This happens for example in the so
called $\nu$-MSSM (SUSY see-saw mechanism) where
off-diagonal elements in the slepton  mass matrix  are
induced by the running of the parameters from the heavy
right-handed neutrino scale to the electroweak symmetry
breaking scale.

The effects are particularly enhanced at large $\tan\beta$
and if the SUSY spectrum is beyond the TeV range LFV
$\tau$-decays like $\tau \to \mu\eta$ and  $\tau \to
\mu\gamma$ are pushed near the experimental upper limit and
their search can be refined or expected to give a positive
result both  at the LHC and at a super-B factory. At the
same time also lepton flavor conserving processes like the
B-physics channels $B \to \mu^+ \mu^-$, $B \to \tau \nu$
and  $B \to X_s \gamma$ have high sensitivity to the above
scenario and put severe constraints on the parameter space.

At forthcoming LHC experiments the heavy neutral Higgs bosons can be
produced copiously and the decay $(A, H) \to \mu\tau$
detected~\cite{brignole1,moretti}. An analysis at the
future ILC in the $e^+ e^-$ mode was carried out in
Ref.~\cite{kanemura1}. In this paper we  extend this
analysis to explore the potential of a photon collider,
which is known to  be an interesting option of the ILC, in
detecting LFV signal mediated by heavy neutral Higgs bosons
of the MSSM. In~\cite{choi} it was shown that within the
considered scenario the process $\gamma \gamma \to \tau\tau
b \bar{b}$ ($\tau\tau$-fusion) is the principal mechanism
for heavy Higgs production in photon-photon collisions and
that it allows to measure $\tan\beta$ with a precision
which is better than that obtainable at the LHC. In this
work we show  that in $\gamma\gamma$ collisions the
dominant channel for the lepton flavor violating process
$\gamma\gamma \to \mu\tau b\bar{b}$ is that in which the
colliding photons split respectively into a $\mu$-pair and
$\tau$-pair and the two virtual leptons ($\mu\tau$)
annihilate into the Higgs which decay into a $b\bar{b}$,
Fig.~\ref{diagrams}(a). We give both an analytical
approximation of the cross section and a detailed numerical
study of the signal by evaluation of the contributing
diagrams discussing background and the necessary cuts to
isolate the signal. The observability of the signal has
been studied by imposing on the large parameter space the
constraints on the SUSY spectrum obtained by electro-weak
precision observables, direct search, $B$ physics and
experimental upper bounds on the rare $\tau$-LFV decays. We
have considered as reference values of the technical
parameters of the photon collider  those of the TESLA
project assuming an integrated luminosity  ${\cal
O}(200-500)$ fb$^{-1}$/yr and $\sqrt{s_{\gamma\gamma}}=600$
GeV.

Let us  finally summarize the results obtained in this
work: $\bm{i}$) for large values of $\tan\beta$,
($30<\tan\beta<60$) the cross
 of $\gamma \gamma \to
\mu\tau b\bar{b}$ goes from $10^{-2}$ fb up to a few fb for
higgs bosons masses ranging from  $M_{A,H}>150$ up to the
kinematical limit, $600$ GeV; $\bm{ii}$) the heavy neutral
Higgs ($A, H$) are practically degenerate in mass and give
the same contribution to the signal cross section;
$\bm{iii}$) the $\mu$ and $\tau$ leptons are emitted
preferentially back-to-back and are characterized by high
energy and low transverse momentum. A cut on the energy up
to $50$ GeV can be safely applied without affecting very
much the cross-section while on the contrary  a cut on
transverse momentum would decrease the signal cross-section
drastically. The b-tagged jets from the Higgs decay have
invariant mass distributions which are peaked at the Higgs
mass. A cut on the invariant mass is sufficient to reduce
significantly the background; $\bm{iv}$) low energy
constraints put further conditions on the observability of
the signal: the parameter space is restricted to those
regions which are allowed by the low energy constraint and
there we look for points were the signal cross section are
in the range $10^{-1} -10^0$ fb or $\sigma > 10^{-1}$,
giving up to $250$ events/yr for the high luminosity option and
up to $100$ events/yr for the low luminosity case;
$\bm{v})$ once the low energy constraints are applied to the parameter
space we find that the lepton flavor violating  signal can be probed for
masses of the heavy neutral Higgs bosons
$A,H$ from $300$ GeV up to the kinematical limit $\simeq 600$ GeV
for 30$\leq\tan\beta\leq$60.

\section*{ACKNOWLEDGMENTS}
The authors would like to acknowledge P.~Paradisi for collaboration in the early stages
of this work and for many useful discussions.

\end{document}